\documentclass[12pt, titlepage]{article}
\usepackage{amsmath,amssymb,amsthm,amsfonts}
\usepackage[mathscr]{eucal}
\usepackage{fullpage}
\usepackage{graphicx, subfigure}
\usepackage{subfig}
\usepackage{epstopdf}
\usepackage{lscape}
\usepackage{setspace}
\usepackage{epsfig}
\usepackage[round]{natbib}
\usepackage{multirow}
\usepackage{color}
\usepackage[latin1]{inputenc}
\usepackage{bbm}
\usepackage{mathtools}
\usepackage{appendix}
\usepackage{siunitx}
\usepackage{booktabs}
\usepackage{array}
\usepackage{multirow}
\usepackage{rotating}
\usepackage{tabularx}
\usepackage{eurosym}
\usepackage{appendix}
\usepackage{bm}

\numberwithin{equation}{section} 

\newtheorem{proposition}{Proposition}[section]
\newtheorem{definition}{Definition}[section]

\def\XX{\boldsymbol{X}}

\def\VaR{\text{VaR}}
\def\ES{\text{ES}}

\DeclareMathOperator*{\argmax}{arg\,max}

\begin{document}

\author{ M. DORIA  \and  \textit{\ 
Credit Suisse Services AG, Quantitative Analysis \& Technology }  \and  {CCM
Credit Model Solutions, CQCB 3.}  \\
 E. LUCIANO\thanks{
Corresponding author: Elisa Luciano, ESOMAS Department and Collegio Carlo Alberto, Universit\'a di Torino. Email: elisa.luciano@unito.it}\and \textit{\ ESOMAS Department and Collegio Carlo Alberto, }Universit\'a di Torino \and
P. SEMERARO
\and  \textit{\ 
Department of Mathematical Sciences G. Lagrange,}  \and  {Politecnico di
Torino.}}
\title{Machine Learning in the default prediction of
credit portfolios: the extra advantage. }
\maketitle

\begin{abstract}
This paper studies the consequences of using Machine Learning (ML) to
predict default in portfolios of credits. We use a large pool of credit card
holders to show that ML algorithms, by their very nature, permit to properly
embed the dependence of obligors' default. Traditional methods like the
logistic regression underestimate the joint losses and the ensuing risk
measures: Value at Risk ($\VaR$) and Expected Shortfall (ES). The result
obtains using $\VaR$ and $\ES$ bounds robust with respect to the model used
for joint default modelling and adopting a Bernoulli mixture approach, which
already encompasses traditional structural and reduced form models. We
consider the superior ability to handle joint - on top of single - defaults
the extra and true advantage of ML in default prediction.
\end{abstract}

\noindent \textbf{keywords}: 
Finance; Risk analysis; Bernoulli mixture model; ML methods; credit cards.

\section{Introduction}

The prediction of  default both of single and groups of
obligors remains one of the important applications of statistical learning and operational research. The
prediction for single obligors is traditionally performed using a
statistical tool, the first order Logistic Regression (LR). Machine Learning
(ML) techniques have only recently been
considered as alternatives. Most of the literature so far has searched for
the most accurate ML method for single obligor defaults.

This paper studies the consequences of using ML to predict default in large
portfolios of credits. We use a pool of credit card holders because
\noindent their joint - and not only single - default matters to the card
issuing company.

For each obligor the credit card data gives us a snapshot of a number of
covariates at a specific point in time and the default indicator one point
in time later. The covariates include socio-economic indicators as well as
present and past bill and payment values. Default is assumed to depend on
those covariates. As a preliminary analysis, ML and  first order LR are used to describe
the - respectively non linear and linear - relationship between the
covariates and the probability of default for each obligor. As already shown
in the literature, the traditional fit measures (accuracy, sensitivity, ROC,
AUC, F1) assign a superiority to the ML techniques.

The very contribution of the paper consists in showing that the superiority
still holds when predicting the loss deriving from the entire portfolio,
through its risk measures, $\VaR$ and $\ES$. A key point of this result is the availability of $\VaR$ bounds, independent of the joint default distribution. Intuitively, superiority depends on the very
nature of ML algorithms, which permit to embed the linear and non linear
dependence of obligors' covariates and therefore of their default.
Traditional methods like the logistic regression capture only the linear
dependence (or a polynomial approximation of the true dependence),
underestimate the losses and the ensuing risk measures.

We show
that capturing correctly the dependence is  not only as important as
fitting appropriately the marginal default probability, as one could expect,
but also that is more important than fitting the higher order moments of the
joint loss. The result is the greater the higher is the portfolio dimension,
namely the number of obligors it contains. It holds both per se and when
comparing the LR/ML risk measures with the $\VaR$ and ES bounds.

Without loss of generality, together with LR and ML, we use a Bernoulli
mixture approach for joint default modelling. We assume that,
conditional on the covariates, defaults in the portfolio can be represented
as independent Bernoulli variables. The latter model enjoys analytical
tractability, is easy to simulate also for high-dimensional portfolios like
the credit card one, and encompasses traditional
structural and reduced-form models. Last but not least, thanks to the Bernoulli set up, by using it  in
conjunction with ML - for the first time, up to our knowledge, we also show that it keeps the modelling complexity the
same between traditional estimation approaches like the LR and newer ones,
like ML. It will therefore be our workhorse model.

The distribution of individual defaults fully determines the estimate of the credit card
issuer's loss, with no need for additional assumptions such as copulas. The Bernoulli approach we use can be parametrical or not. The former
is highly recommended for large portfolios, while for small portfolios we
 also use a non-parametric distribution of the loss. We show that the
latter is unfeasible for simulating risk measures beyond a very limited
number of obligors. The use of the Bernoulli parametric distribution will
not come at a disadvantage, independently of the fit of the overall
distributions, thanks to the way we estimate its moments.

Overall, then we point to the extra-advantage of ML, using a mixture approach that has a number of advantages including the fact of putting ML and LR on the same level playing field, in terms of modelling complexity. We show that the superiority depends on the first two moments of the loss and is robust respect to  using parametric or non parametric losses.

The paper is organized as follows: Section \ref{back} illustrates the
background literature in univariate ML default prediction, especially the
credit card one. Section \ref{not} sets up the multivariate model
consistently with the Bernoulli mixing approach. Section \ref{RD} specifies
the research design and reviews both the LR/ML techniques used in the paper
and their single-obligor fit measures. Section \ref{DD} illustrates the
credit card data and its basic features, together with the  robustness procedures
applied to it for probability fitting. Section \ref{Res} explains how we fit
the individual and joint default probabilities in the LR and in the three ML
cases, and compares their fit. Section \ref{VaR} studies the corresponding
measures of the high quantiles of the portfolio loss distribution with
respect to their bounds. The Appendices contain
some technical specifications.

\bigskip

\section{Background literature\label{back}}

For single obligors, there is a well established literature which studies
the assignment of a credit scoring and a default probability. Classic
instruments for probability prediction are LR and discriminant analysis.
More recently, ML and Artificial Intelligence (AI) algorithms have been
exploited. First, researchers have used individual classifiers, that include
for instance K-Nearest Neighbors, naive Bayesian, Neural Networks and
classification trees (see for instance \cite{yeh2009comparisons}). Later on,
they have adopted also ensemble methods, that include Random Forest and
Support Vector Machines (see for instance \cite{chen2021predicting} ).
Lately, some Authors have used Boosting techniques, leading to the
development of the Ada Boost and GBoost or XGBoost techniques (see for
instance \cite{chen2021predicting}). For an explanation of their main
features and differences among individual we refer to \cite{kenett2021modern}, while for a
survey of the applications of AI other than ML - for instance Neural
Networks and Deep Neural Networks - in credit risk we suggest \cite{gunnarsson2021dep}.

We adopt an individual,  ensemble and  boosted ML algorithms, together with
the first order LR method. In particular, we use K-Nearest Neighbors, Random
Forest and Ada Boost, because all of them are consistent, namely their
distribution shrinks to the true parameter value when the sample dimension
grows (see \cite{devroye1994strong}, \cite{scornet2015consistency}, \cite%
{bartlett2006adaboost}). We describe their specifications below.

K-Nearest Neighbors in the form of closest neighbour was used for the first
time in the credit risk domain by \cite{chatterjee1970nonparametric}, who
pointed at its non-parametric, simple applicability to credit scoring, and
computed the probability of misclassification. It was then used also for the
computation of the probability of default. Random Forest, with the specific
purpose of computing the default probability, appears in \cite%
{kruppa2013consumer}. \cite{tsai2014comparative} presented bagging and
Boosting methods for different ensemble techniques and analyzed their
performance in bankruptcy prediction.

The literature has had difficulty in assessing whether one of the ML
approaches performs better than the others, while it is quite unanimous in
concluding that they perform better than the first order LR, in
single-default probability prediction. ML is found to outperform LR, whether
it uses individual, ensemble or boosted technologies.

Using individual algorithms, \cite{yeh2009comparisons} compare the ability
to predict the default probability of different individual algorithms and
LR. They conclude that Neural Networks do outperform LR.

Using both individual and ensemble methods, \cite{lessmann2015benchmarking}
compare 41 ML criteria and the LR. They point at the overall superiority of
ML techniques with respect to the LR and find that the best performer is the
Random Forest. Dumitrescu et al (2021) use a penalized LR with data coming
from decision trees to make the LR results comparable with respect to
ensemble methods.

Using ensemble and boosted methods, \cite{wang2018personal} find that GBoost
outperforms Random Forests without Boosting.

\cite{chen2021predicting}, who consider logistic methods, individual and
ensemble ones, show that there is no unique best predictor among the ML ones.

So, the conclusion we feel like taking from recent advances in OR is that
first-order LR is outperformed by ML, at the single obligor level.

However, we have mixed results on the relative performance of different ML
forecasting methods, letting aside LR. The reason is that the performance is
not uniform over measurement tools and data.

In Section \ref{Res} we show that, as known in the ML literature, the
appropriateness of each method for single default prediction really depends
on whether the researcher is more interested in false negative, false
positive, or combinations of the two. These features are measured by
selected metrics, such as precision, sensitivity, specificity, the area
below the so-called ROC curve, and the so-called F1-score, that we
illustrate below.

The data on which the ability of the different ML prediction approaches is
tested is also important for the result. Below, we examine a publicly
available dataset on Taiwanese credit card holders.

The same dataset, which dates back to 2005 and is publicly available from
Kaggle, has already been examined with selected ML and AI algorithms,
including K-Nearest Neighbors, LR and Neural Networks, in \cite%
{yeh2009comparisons}. The authors do not use the F1-score or the area under
the ROC curve. They test the predictive accuracy by regressing the real on
the forecasted probability, where the real probability is produced using a
"sorting smoothing method". They reach the conclusion that the artificial
Neural Network outperforms other individual methods. We depart from their
conclusions, when examining the predictive accuracy of ML methods for single
defaults, because the set of ML algorithms we use is different from their
own, we do not use AI, and therefore Neural Networks, and because the
metrics for assessing the performance is not a regression of their type. We
use the standard metrics of ML mentioned above, in particular the F1-score
and the area under the ROC curve.

Other studies that use ML to predict default in credit cards are \cite%
{bellotti2009support}, \cite{yu2010support}, \cite{hamori2018ensemble}, \cite%
{mahbobi2021credit}. \cite{bellotti2009support} use a different credit-card
dataset, namely a proprietary dataset of 25000 credit card clients of a
private institution, with contracts opened in March 2004. They show that on
their data support vector machines, another ML approach, which we do not
use, is superior to the traditional approaches they adopt, namely LR and
discriminant analysis. Bellotti and Crook focus on the analysis of the
covariates that are most important in driving defaults. We are not
interested in understanding the importance of the single covariates in order
to determine the default of single obligors. We are interested in their
dependence because it is what determines joint defaults.

Yu et al. balance a British credit card dataset - which per se, in default
data, is unbalanced, because usually there is a greater occurrence of no
defaults - and establish that Support Vector Machines (SVM) outperform LR and
quadratic discriminant analysis. Hamori et al. use the University of
California at Irvine dataset and reach the same conclusion as Yu et al.,
without rebalacing the dataset. They use ensemble, Neural Network and Deep
Neural Networks. Mahbobi et al. discuss exactly the issue of dataset
rebalancing versus over and undersampling, on credit card data, in
conjunction with those methodologies, such as SVM (not used here) which are
more sensitive to unbalanced data. We show below, using SMOTE, a technique
which is also in Mahbobi, that on our dataset the estimates are robust to
unbalancing vs balancing the dataset.

Together with Yu et al., the only paper, up to our knowledge, which includes
in the LR also higher order and interaction terms, and compares their
predictive accuracy on individual defaults, is \cite{dumitrescu2022machine}.
In spite of the Authors including polynomial terms to improve the fit of the
LR, the latter remains less accurate than the ML techniques in single
obligor prediction. The intuitive reason why also the inclusion of higher
order terms in the LR does not make it as good in prediction as ML is that,
in any case, polynomial terms provide an approximation to the actual,
non-linear relationship among the covariates, while increasing the number of
parameters to estimate. A paper which testifies that on a different, but
huge dataset of mortgages (120 million) is \cite{sadhwani2021deep}: they
recognize that, being ML entirely dictated by the data themselves, it
minimizes the model mis-specification and the bias of the variable weights'
estimates inherent in - even higher order - LR. That is why below we stop to
the first order LR.

We build on the literature described so far and we take as given -but verify-  that ML
outperform LR to classify single defaults, on credit cards. We make a step
forward and empirically investigate the consequences on the risk of - even
high dimensional - portfolios.

\section{\protect\bigskip Modelling dependent defaults}

\label{not}

To model the occurrence of losses in portfolios, we use a mixture model,
which means that, once we account for the dependence of single defaults on a
set of common covariates, through LR or ML, single defaults are the
realizations of the mixing distribution in an exchangeable Bernoulli mixture
model. This approach - as we recall in Section \ref{EBM} - has the advantage
of providing the probability of any number of joint defaults in quasi-closed
form and to give a one-to-one mapping between the moments of the mixing
variable and the unconditional probabilities of any number of defaults.
Also, it is the counterpart of any structural threshold model in which
defaults are conditionally independent given the value of some common
factors, as proved in \cite{frey2001modelling} and re-assessed in \cite%
{bacsouglu2018efficient}. This means that its adoption can occur without
loss of generality, from the point of view of the economic interpretation.
From the numerical point of view, a  mixture representation
lends itself to easier Monte Carlo simulation and simpler asymptotic results
- when the dimension of the credit portfolio becomes large - than alternative ones (see \cite{mcneil2005quantitative}).

A standard reference for mixture models, and the particular one we use, the
Bernoulli one, is \cite{mcneil2005quantitative}. We differ from the
literature on Bernoulli mixture models in the formulation of the link
between the marginal and joint unconditional default probabilities and the
covariates, which here relies on ML.

Let the random vector $\boldsymbol{Y}=(Y_{1},\ldots ,Y_{d})$ be the vector
of default indicators of a set of $d$ obligors or credit card owners over a
fix time horizon $T$, that in our case is one month. Let $P=(w_{1},\ldots
,w_{d})$ be the percentage weights which represent a credit risk portfolio
at time $T$ associated to the $d$ obligors, where $w_{i}\in (0,1]$ and $%
\sum_{i=1}^{d}w_{i}=1$. To model the loss of the portfolio $P$, we consider
the sum of the percentage individual losses $L$, given by:
\begin{equation}
L=\sum_{i=1}^{d}w_{i}Y_{i},
\end{equation}%
In this paper we consider the case $w_{i}=\frac{1}{d},\,\,\,i=1,\ldots ,d$.
The extension to unequal weights can be done numerically or by simulation. A
relevant variable is the number of defaults,
\begin{equation}
S=\sum_{i=1}^{d}Y_{i},
\end{equation}%
that fully characterizes the loss in the case of equal weights, where $L=S/d$%
.

To represent $\boldsymbol{Y}$ and $S$ we use a Bernoulli mixture model,
according to the following

\begin{definition}
\label{BMM} Given some $n<d$ and a $n-$dimensional random vector $\bm{\psi}%
=(\psi_1,\dots,\psi_n)$, the random vector $\bm{Y}=(Y_1,\dots,Y_d)^{\prime }$
follows a Bernoulli mixture model with factor vector $\bm{\psi}$, if there
are functions $q_i :\mathbb{R}^p\rightarrow [0,1]\quad 1\leq i\leq d$, such
that conditional on $\bm{\psi}$ the default indicator $\bm{Y}$ is a vector
of independent Bernoulli random variables with $\mathbb{P}(Y_i =1 |\bm{\psi}%
)=q_i(\bm{\psi})$.
\end{definition}

In a mixture model the default probability of an obligor is assumed to
depend on a set of (often unobservable) economic factors $\boldsymbol{\psi }$%
: given $\boldsymbol{\psi }$ defaults of different obligors are independent.

\subsection{Exchangeable Bernoulli mixture model}

\label{EBM} { Usually, $\boldsymbol{\psi }$ is a latent variable and has a
factor structure to model common and idiosyncratic factors that affect
defaults. Here we depart from this model and we assume that the
distributions of individual defaults, $q_{i}(\bm{\psi})$ in Definition \ref%
{BMM} are the same  function of the same set of random covariates $\boldsymbol{X}$%
, thus we write $q(\boldsymbol{X})$. We therefore assume that there is
only one distribution of default probabilities, i.e. $q(\boldsymbol{X}%
)=h(\boldsymbol{X})$ for all $i$. We interpret this as modelling a
"homogeneous" group of obligors. 

Formally, {since default probabilities are conditionally independent given a
single common mixing variable $Q=h(\boldsymbol{X})$, that represents the
distribution of the individual default probabilities of obligors, we assume
that the Bernoulli mixture model is exchangeable.}}

\begin{definition}
Given a random variable $Q$, the random vector $\bm{Y}=(Y_1,\dots,Y_d)^{%
\prime }$ follows an exchangeable Bernoulli mixture model with mixing
variable $Q$ with support on $[0,1]$, if conditional on $Q$ the default
indicator $\bm{Y}$ is a vector of independent Bernoulli random variables
with $\mathbb{P}(Y_i =1 |Q )=Q$.
\end{definition}

Since $Q=Q_{h}=h(\boldsymbol{X})$ is function of a vector of { random}
covariates $\boldsymbol{X}$, the realizations of $Q_{h}$ are functions of
the realizations of $\boldsymbol{X}$, $q=h(\boldsymbol{x}).$ The
conditional default probability is
\begin{equation}
\mathbb{P}(Y_{i}=1|Q_{h}=h(\boldsymbol{x}))=h(\boldsymbol{x})
\label{condi}
\end{equation}%
The unconditional marginal default probability becomes
\begin{equation}
\mathbb{P}(Y_{i}=1)=\int_{0}^{1}qdG_{h}(q),
\end{equation}%
where $G_{h}(q)$ is the distribution of $Q_{h}$, and the unconditional
probability mass function (pmf) $p_{\boldsymbol{Y}}(\boldsymbol{y})$ of $%
\bm{Y}$ becomes:
\begin{equation}
p_{\boldsymbol{Y}}(\bm{y})=\mathbb{P}(\bm{Y}=\bm{y})=\int_{0}^{1}q^{\sum%
\limits_{i=1}^{d}y_{i}}(1-q)^{d-\sum\limits_{i=1}^{d}y_{i}}dG_{h}(q).
\end{equation}%
%
%
We introduce the following simple notation for the cross moments of $%
\boldsymbol{Y}$:
\begin{equation}
\pi _{k}=E[Y_{i_{1}}\cdots Y_{i_{k}}],\,\,\,\text{$\{i_{1},\dots
,i_{k}\}\subset \{1,\dots ,d\}$\thinspace \thinspace \thinspace\ $1\leq
k\leq d$}.  \label{cm}
\end{equation}%
Notice that $\pi _{1}=P(Y_{i}=1)$ is the marginal default probability; we
call it $p:=\pi _{1}$, while $\pi
_{2}=E[Y_{i_{1}}Y_{i_{2}}],\,\,\,\{i_{1},i_{2}\}\subset \{1,\dots ,d\}$; $%
\pi _{k}$, the $k-$th order joint default probability is the probability
that an arbitrary selected subgroup of $k$ obligors defaults in $[0,T]$. We
can easily compute the following relevant quantities:

\begin{equation}
\begin{split}
E[Y_{i}]& =p,\,\,\,\ Var(Y_{i})=p(1-p) \\
Cov(Y_{i},Y_{j})& =p(1-p),\,\,\,\rho =\rho (Y_{i},Y_{j})=\frac{\pi _{2}-p^{2}%
}{p(1-p)}\qquad i\neq j.
\end{split}
\label{corr}
\end{equation}%

{We remark that $\rho$ is the correlation among defaults, that is a function of the marginal default probability and the second order moment of the univariate mixing variable. Dependence among covariates is captured by the conditional default probabilities, i.e. by the distribution of individual defaults, that is the distribution of $Q$.}

The unconditional distribution $S$ of the number of defaults can be computed
in terms of the $\pi _{k}$s in the following way (see \cite%
{frey2001modelling}):
\begin{equation}
p_{S}(k)=\mathbb{P}(S=k)=\sum_{i=0}^{d-k}(-1)^{i}\frac{d!}{i!k!(d-k-i)!}\pi
_{k+i}.  \label{pippo}
\end{equation}%
It is entirely determined by the joint distributions of the default
indicators. This is a consequence of the exchangeability of the vector $%
\boldsymbol{Y}$. In fact if $\boldsymbol{Y}$ is exchangeable there is a one
to one correspondence between the distribution of the number of defaults and
the joint distribution of defaults (\cite{fontana2021model}). The
unconditional distribution $p_{S}(k)$ of the number of defaults $S$ becomes:
\begin{equation}
p_{S}(k)=\binom{d}{k}\int_{0}^{1}q^{k}(1-q)^{d-k}dG(q).  \label{E1}
\end{equation}%
In an exchangeable Bernoulli mixture model it can be proved that the cross
moments of $\boldsymbol{Y}$ are the moments of the mixing variable $Q_{h}$:
\begin{equation}
\pi _{k}=\mathbb{E}[Q_{h}^{k}],  \label{E2}
\end{equation}%
In particular
\begin{eqnarray*}
\pi _{1} &=&\mathbb{E}[Q_{h}]=p, \\
Cov(Y_{i},Y_{j}) &=&Var(Q_{h}),.i\neq j
\end{eqnarray*}
Comparing \eqref{E2} with \eqref{cm} and \eqref{pippo} it follows that the
moments of the mixing variable $Q_{h}$ completely determine the joint
distribution of defaults and consequently the distribution of the number of
defaults. Therefore, using the sample moments of $Q_{h}$, that are a
function of the observable covariates, we can in principle estimate the
distribution of $S$. This can be done in practice for low dimensions, since
the non parametrical distribution in \eqref{pippo} exhibits numerical
problems from $m\sim 30$ onwards. The specification of a parametrical
distribution for $Q_{h}$ allows us to work also in high dimensions.

 Because the
distribution of $Q_{h}$ depends on the distribution of the covariates, we do
not have any information about its parametrical form. {The model risk
associated to the parametrical form of the distribution of $Q_{h}$ has been
discussed in \cite{mcneil2005quantitative}, Section 8.4.6. Proposition 8.16
in \cite{mcneil2005quantitative} proves that the tail distribution of the
number of defaults is determined by the tail distribution of $Q_{h}$.
Nevertheless, \cite{mcneil2005quantitative} conclude that, if the first two
moments are fixed, this seems to be significant only after the $0.99$
quantile of the distribution. When excluding the extreme percentile, the
more relevant quantities or the total number of default distribution are the
marginal default probability and the default correlation, $p$ and $\rho $.
For this reason, we consider the particular parametrical form of the mixing
variable less important than the estimates of $p$ and $\rho $ and we opt for
} a beta distribution with parameters $a$ and $b$ for $Q_{h}$, i.e. $Q_h\sim
\beta (a,b)$. This gives rise to an exchangeable Bernoulli mixture model
frequently used in practice, i.e. the beta mixing-model.
If $Q_h\sim \beta (a,b)$ then the number of defaults $S$ follows a so-called
beta-binomial distribution of parameters $d$, $a$ and $b$ (see \cite%
{mcneil2005quantitative}), where we recall that $d$ is the dimension of $%
\boldsymbol{Y}$ or the number of obligors. The beta and the beta-binomial
distributions are recalled in Appendix \ref{beta}. 

\section{Research design}

\label{RD}

The classical model specification for the function $h$ that estimates the
individual default probabilities $q_{i}$, and consequently the moments of $%
Q_{h}$, is a first order LR. We are to compare it with different ML
techniques as model specifications $h$ for $Q_{h}$.
{We decide to consider three of
the most popular ML methods to empirically support the idea that their impact on the
loss is similar and differs from the impact of LR}. Comparing the performance of different ML techniques to
predict single defaults is out of our aim.  We have a total of four
models: Random Forest (RF), Ada Boost (AB) and K-Nearest
Neighbors (KNN), reviewed in Section \ref{ML}. To each one of them a
different corresponds a different $h$ function, $h=$ LR, RF, AB, KNN. %

The ultimate goal is to measure the risk of the loss of a multivariate
portfolio associated to the choice of a ML technique instead of the
classical LR to estimate $Q_{h}$. We therefore consider the risk associated
to the corresponding distribution of the number of defaults. As a measure of
risk we consider the $\VaR$, that is a very popular measure of risk among
financial institutions, due to regulatory requirements, and the $\ES$, that
is a coherent measure of risk.\footnote{%
See \cite{artzner1999coherent} for the axioms that characterize a coherent
risk measure. See \cite{acerbi2002coherence} for a proof of the coherence of
the ES.} We recall their definition for a general random variable $Y$.
\begin{definition}
Let $Y$ be a random variable representing a loss with finite mean. Then the $%
\text{VaR}_{\alpha }$ at level $\alpha $ is defined by
\begin{equation*}
\text{VaR}_{\alpha}(Y)=\inf \{y\in \mathbb{R}:P(Y\leq y)\geq \alpha \}
\end{equation*}%
and the expected shortfall at level $\alpha $ is defined by
\begin{equation*}
\text{ES}_{\alpha }(Y)=\frac{1}{1-\alpha }(E[Y;Y\geq \text{VaR}_{\alpha
}(Y)]+\text{VaR}(Y)(1-\alpha -P(Y\geq \text{VaR}(Y)))).
\end{equation*}
\end{definition}
{As known, the $\VaR$ is the $\alpha $ quantile of the distribution of $S$,
while the }$\ES$ is the expectation of the losses beyond the $\VaR$ threshold,
and is usually greater than the $\VaR$ itself. At most the two coincide{. }

Both the $\VaR$ and the $\ES$ for exchangeable Bernoulli variables have analytical
bounds, which depend on the marginal default probability $p$. Indeed, let $%
\mathcal{S}_{d}(p)$ the class of all distributions of sums of exchangeable $%
d $-dimensional sums of Bernoulli distributions with mean $p$. We refer to
\cite{fontana2021model} for the proof of the following result.

\begin{proposition}
\label{Varbound} Let us consider the class $\mathcal{S}_d(p)$. Let $j_1^M$
be the largest integer smaller than $pd$, $j_2^m$ be the smallest integer
greater than pd and $j_1^p=\frac{(p-(1-\alpha))d}{\alpha}$.

\begin{enumerate}
\item If $p<1-\alpha $, $\min_{S\in \mathcal{S}_{d}(p)}\text{\VaR}_{\alpha
}(S)=0$ and $\max_{S\in \mathcal{S}_{d}(p)}\text{VaR}_{\alpha }(S)=[\frac{pd%
}{1-\alpha }]$ if $\frac{pd}{1-\alpha }$ is not integer and $\max_{S\in
\mathcal{S}_{d}(p)}\text{VaR}_{\alpha }(S)=\frac{pd}{1-\alpha }-1$ if it is
integer.

\item If $1-\alpha \leq p\leq 1-\alpha +\frac{\alpha }{d}j_{1}^{M}$, $%
\min_{S\in \mathcal{S}_{d}(p)}\text{VaR}_{\alpha }(S)=j_{1}^{\ast }$, where $%
j_{1}^{\ast }$ is the smallest integer greater or equal to $j_{1}^{p}$ and $%
\max_{S\in \mathcal{S}_{d}(p)}\text{VaR}_{\alpha }(S)=d$.

\item If $p>1-\alpha +\frac{\alpha }{d}j_{1}^{M}$, $\min_{S\in \mathcal{S}%
_{d}(p)}\text{VaR}_{\alpha }(S)=j_{2}^{m}=j_{1}^{M}+1$ and $\max_{S\in
\mathcal{S}_{d}(p)}\text{VaR}_{\alpha }(S)=d$. In this case, if $pd$ is
integer $j_{1}^{M}+1=pd$.
\end{enumerate}

Let $ES_{\alpha }(S_{d})$ be its expected shortfall. Then
\begin{equation*}
\min_{S\in \mathcal{S}_{d}(p)}\text{$\VaR$}_{\alpha }(S)\leq \text{$\ES$}%
_{\alpha }(S_{d})\leq d.
\end{equation*}
\end{proposition}

The bounds for $\ES$ are weak and trivial. Nevertheless, at least
in some cases, they cannot be improved (see \cite{fontana2021model}). For
this reason, we consider the $\VaR$ bounds. {Different choices of the
function $h$, i.e. of the LR/ML technique, give different estimates of }$p$,
and therefore different $\VaR$ bounds, which we will compare below to assess
whether one or more among the LR/ML techniques tends to underestimate the
bounds.

Apart from the $\VaR$ bounds, to assess whether one or more among the LR/ML
techniques tends to underestimate the $\VaR$ itself we  simulate using the parametrical (beta-binomial)
distribution or a non parametrical version of it and, in that case, whether
we need only some moments (say, the first two) or all of them.
The computational limit for the non parametrical distribution of $S$ is
around 25.

 As for the needed moments, we expect the $\VaR$ itself to depend
on {the first two moments of the mixing variable, since} we established
above that {the latter affect the tail of the number of defaults. Different
choices of the function $h$ give different estimates of these moments, and
different }$\VaR$s. In particular, they give different first two moments:
not only \ different $p$s, but also {different second order moments }$\pi
_{2}{.}$ The latter, {combined with $p$, give different correlations $\rho ,$
according to (\ref{corr}). The correlation $\rho $ is decreasing in $p$ and
increasing in $\pi _{2}$. Since $p$ enters in the portfolio $\VaR$ both
directly and via the correlation }$\rho $, it is {\ worth investigating its
role in determining the relative performance of ML approaches vs the LR one.
Higher marginal default probabilities should increase the
risk of the portfolio, and therefore the $\VaR$. Also, because  higher
default probabilities lead to lower correlations, {that should decrease the }$%
{\VaR}$ as a result of greater diversification.}

To determine the effect of $p$ we will study the $\text{VaR}$ bounds. To determine the effect of $p$ and $\pi_2$, or equivalently $p$ and $\rho$, we will study the $\VaR$ estimated using the parametrical distribution of the mixing variable. To determine the effect of all moments we will study the VaR using the nonparametrical distribution.

We expect this to confirm
that the higher moments do not matter for risk measures. This approach can
be adopted for low dimension portfolios, for which we have both the
parametrical and non-parametrical distribution of the number of defaults; we
choose $d=25$. For greater portfolios, we can use only the parametrical one.

The research design is put in pratice according to the following operational
steps, performed for each choice of $h$ = LR, RF, AB, KNN.

\begin{enumerate}
\item We estimate $q_j^h=h(\boldsymbol{x}_j)$, where $\boldsymbol{x}_j$ is
the $m$-dimensional vector of  realizations of $\XX$, and find a sample of
estimated conditional default probabilities $\hat{\boldsymbol{q}}^h=(\hat{q}%
^h_1,\ldots, \hat{q}^h_n)$;

\item we compute the marginal default probability $p$ and the
equicorrelation among default indicators $\rho $.

\item we estimate the parametrical beta-binomial distribution of the number
of default $S$ by moments matching, using the first two moments of $Q_{h}$, $%
p$ and $\pi _{2}$;

\item We consider a low-dimension $d=25$ portfolio and:

\begin{enumerate}
\item we analyze the effect of $p$ on the $\VaR$, by using the analytical
bounds provided in Proposition \ref{Varbound};

\item we empirically evaluate the effect of $p$ and $\pi _{2}$ on $\VaR$ and $%
\ES$, by computing the $\VaR$ and $\ES$ for the beta-binomial distribution of
the loss;

\item we estimate the non parametrical distribution of $S$ using equation %
\ref{pippo}, that uses all the sample moments to measure the impact of
higher order moments with respect to the correlation. We use the
Kullback-Leibler (KL) distance to compare the non parametrical and
parametrical distributions of the loss and we choose the best model for the
high-dimension portfolio analysis.
\end{enumerate}

\item We consider a high-dimension $d=6000$ portfolio and:

\begin{enumerate}
\item we analyze the effect of $p$ on the $\VaR$, by using the analytical
bounds provided in Proposition \ref{Varbound};

\item we analyze the effect of the first two moments $p$ and $\pi _{2}$ on
the risk of the aggregate loss, by computing the $\VaR$ and $\ES$ of the
beta-binomial distribution of the loss.
\end{enumerate}
\end{enumerate}


%

\subsection{Classification techniques}

\label{ML} This section recalls the four different classification methods
considered: Logistic Regression, Random Forest, AdaBoost and K-Nearest
Neighbors. It also recalls the metrics used to compare their performance.

\subsubsection{Logistic Regression}

The Logistic regression (LR) is a generalized linear model, where the
individual conditional responses $Y_{1},\ldots ,Y_{n}$ are independent and
identically distributed. In the LR model the conditional probabilities are
given by:
\begin{equation}
q_{i}:=\mathbb{P}(Y_{i}=1|Q_{h}=h(\boldsymbol{x}_{i}))=\frac{e^{\boldsymbol{\beta} ^{T}{%
\boldsymbol{x}_{i}}}}{1+e^{\boldsymbol{\beta}^{T}{\boldsymbol{x}_{i}}}}=\frac{1}{%
1+e^{-\boldsymbol{\beta}^{T}{\boldsymbol{x}_{i}}}}.  \label{probsLR}
\end{equation}%
Solving the previous equation for the exponential we gain a better insight
into the meaning of vector parameter $\beta $:
\begin{equation*}
e^{\boldsymbol{\beta} ^{T}{\boldsymbol{x}}_{i}}=\frac{q_{i}}{1-q_{i}}.
\end{equation*}%
\newline
Then taking the logarithms of both sides, we obtain
\begin{equation*}
\log\bigg(\frac{q_{i}}{1-q_{i}}\bigg)=\boldsymbol{\beta}^{T}\boldsymbol{x}_{i}.
\end{equation*}%
The ratio $q_{i}/(1-q_{i})$ provides us with equivalent information in terms
of odds. The logarithm of the odds is known as a logit function.\newline
In order to find a way to fit the regression coefficients against a set of
observations ${\boldsymbol{x}_{i}}$ and $y_{i}\in \{0,1\}$, $i=1,\dots ,n$
the maximum-likelihood method is used.\newline
The target variable $Y_{i}$, which can take values $y_{i}$ in the set $%
\{0,1\}$, may be regarded as the realization of a Bernoulli variable, whose
density probability function is
\begin{equation*}
\mathbb{P}(Y_{i}=y_{i}|Q_{h}=q_{i})=q_{i}^{y_{i}}(1-q_{i})^{1-y_{i}}.
\end{equation*}%
\newline
From the expression \eqref{probsLR} for the probability $q_{i}$ we know that
$q_{i}$ depends on the regressors $\mathbf{x_{i}}$ and the vector of
parameters $\mathbf{\beta }$.\newline
Assuming independence of errors, observations are independent as well, and
the likelihood function is just the product of individual probabilities:
\begin{equation}
L=\prod\limits_{i=1}^{n}p_{i}^{y_{i}}(1-p_{i})^{1-y_{i}}=\prod%
\limits_{i=1}^{n}\bigg(\frac{1}{1+e^{{-\boldsymbol{\beta} ^{T}\boldsymbol{x}_{i}}}}\bigg)%
^{y_{i}}\bigg(\frac{e^{{-\boldsymbol{\beta} ^{T}\boldsymbol{x}_{i}}}}{1+e^{{-\boldsymbol{\beta} ^{T}%
\boldsymbol{x}_{i}}}}\bigg)^{1-y_{i}}.
\end{equation}%
The task of maximizing $L$ with respect to the vector of coefficients $%
\mathbf{\beta }$ can be simplified by taking its logarithm:
\begin{equation}
\begin{split}
\mathcal{L}=\log L& =\sum\limits_{i=1}^{n}\log \bigg(\frac{1}{1+e^{\mathbf{%
-\boldsymbol{\beta} ^{T}}\boldsymbol{x}_{i}}}\bigg)y_{i}+\sum\limits_{i=1}^{n}\log \bigg(%
\frac{e^{{-\boldsymbol{\beta} ^{T}\boldsymbol{x}_{i}}}}{1+e^{{-\boldsymbol{\beta} ^{T}\boldsymbol{x}%
_{i}}}}\bigg)(1-y_{i}) \\
& =\sum\limits_{i=1}^{n}y_{i}{\boldsymbol{\beta} ^{T}\boldsymbol{x}_{i}-\log (1+e^{\boldsymbol{\beta}
^{T}\boldsymbol{x}_{i}})}
\end{split}%
\end{equation}%
Thus the goal of the LR is to find the vector of coefficients $\mathbf{\beta
}$ that maximizes this log-likelihood function, namely to solve this
optimization problem:
\begin{equation}
\boldsymbol{\beta} =\argmax_{\boldsymbol{\beta} }\mathcal{L}.
\end{equation}

\subsubsection{K-Nearest Neighbors}

Nearest Neighbor (NN) algorithms are among the simplest of all machine
learning algorithms. A label - \ a realization of default or non-default, in
our case - is attached to any sample realization. A distance is chosen, so
as to separate closest from further "neighbor samples" based on their
distance. The idea behind the NN model is that, by storing all the samples,
one can try to predict the label of any new instance on the basis of the
label of its closest neighbors in the training set. The rationale behind
such a method is that close-by points likely to have the same label.\newline
Let {$\mathcal{X}$, the support of the covariate random vector $\boldsymbol{X%
}$, }be endowed with a metric function $\rho :\mathcal{X}\times \mathcal{X}%
\rightarrow \mathbb{R}$ that returns the distance between any two elements
of $\mathcal{X}$. For example, if $\mathcal{X}=\mathbb{R}^{d},$then $\rho $
can be the Euclidean distance. However, depending on the type of data we
have, there are other popular distance measures, like the Hamming distance,
which computes the distance between binary vectors, the Manhattan distance,
which computes the distance between real vectors using the sum of their
absolute difference and finally the Minkowski distance, which is a
generalization of the previous ones.\newline
The KNN algorithm is as follows: given a positive integer $K$ and a test
observation $\boldsymbol{x}_{0}$, the KNN classifier first identifies the $K$
points in the training data that are closest to $\boldsymbol{x}_{0},$ in
that they belong to a set $\mathcal{N}_{0}$. It then estimates the
conditional probability for $Y_{i}$ to be in class $j,j=0,1,$ as the
fraction of points in $\mathcal{N}_{0}$ whose responses equal $j$:

\begin{equation}
\mathbb{P}(Y_{i}=j\lvert \boldsymbol{X}=\boldsymbol{x}_{0})=\frac{1}{K}%
\sum_{i\in \mathcal{N}_{0}}{I}(y_{i}=j).
\end{equation}%
where ${I}$ is the indicator function.

Finally, KNN applies Bayes rule and classifies the test observation $%
\boldsymbol{x}_{0}$ in the class to whom it has the largest probability of
belonging.\newline

\subsubsection{Random Forest}

A decision tree is a predictor
\begin{equation*}
h:\mathcal{X}\rightarrow \{0,1\},
\end{equation*}%
for the default-no-default or $\{0,1\}$ label associated with an instance $%
\bm{x}_{i}$ obtained by travelling from a so-called root node of a tree to a
so-called leaf. Each node of the root-to-leaf path represents a variable,
and at each node the data or input space is split into two successors based
on the value of that variable or on a predefined set of splitting rules. By
successive splitting, the tree arises, until a specific leaf is reached.
Each leaf contains a specific label.\newline
One of the main advantages of decision trees is that the resulting
classifier is very simple to understand and graphically interpret. The
decision trees suffer from high variance in their results. This means that
if we split the training data into two parts at random, and fit a decision
tree to both halves, the results that we get could be quite different. In
contrast, a procedure with low variance will yield similar results if
applied repeatedly to distinct data subsets; for example, LR tends to have
low variance if the ratio of $n$ (number of observations) and $d$ (number of
features) is moderately large.\newline
Decision trees can be improved by a procedure called bagging, which aims at
reducing the variance of a statistical learning method. Take repeated
samples, say $J$, from the training set, build a separate prediction model
using each bootstrapped training set, by calculating $\hat{h}^{1}(%
\boldsymbol{x}_{i}),\hat{h}^{2}(\boldsymbol{x}_{i}),\dots ,\hat{h}^{J}(%
\boldsymbol{x}_{i})$, and then take a majority vote among all the $J$
predictions. The overall prediction is the most commonly occurring class
among the $J$ predictions.\newline
Random Forests provide an improvement over bagged trees by way of a small
tweak that decorrelates the trees. As in bagging, a number of decision trees
on bootstrapped training samples is built. When building these decision
trees, each time a split in a tree is considered, a random sample of $m$
predictors is chosen from the full set of $d$ predictors. The split is
allowed to use only the $m$ predictors. A fresh sample of $m$ predictors is
taken at each split, typically with $m\approx \sqrt{d}$. \ By so doing, in
building a Random Forests, at each split in the tree, the algorithm is not
even allowed to consider a majority of the available predictors. Random
Forests overcome this problem by forcing each split to consider only a
subset of the predictors. If there is a strong and some weaker predictors,
on average $(d-m)/d$ of the splits will not even consider the strong
predictor, and so other predictors will have more of a chance. We can think
of this process as decorrelating the trees, thereby making the majority vote
among the resulting trees more reliable.\footnote{%
If there is one very strong predictor in the data set, along with a number
of weaker predictors. Then in the collection of bagged trees, most or all of
the trees will use this strong predictor in the top split. Consequently, all
of the bagged trees will look quite similar to each other, and the
predictions from the bagged trees will be highly correlated. Unfortunately,
taking a majority vote among many highly correlated quantities does not lead
to as large of a reduction in variance as taking a majority vote among many
uncorrelated quantities. This means that bagging will not lead to a
substantial reduction in variance over a single tree.\newline
}\newline

\subsubsection{Ada Boost}

The AdaBoost algorithm receives as input a training set of samples $%
S=\{(x_{1},y_{1}),\dots ,(x_{m},y_{m})\}$ where for each $i$, $y_{i}$
depends on $x_{i}$ through some labelling function $f,y_{i}=f(x_{i})$. The
Boosting process proceeds in a sequence of consecutive rounds. At round $t$,
the booster first defines a distribution over the samples in $S$, denoted $%
D^{(t)}$, that is such that $D^{(t)}\in \mathbb{R}_{+}^{m}$ and $%
\sum\limits_{i=1}^{m}D_{i}^{(t)}=1$. Then, the booster passes the
distribution $D^{(t)}$ and the sample $S$ to the "weak learner", in such a
way that the weak learner can construct i.i.d. samples according to $D^{(t)}$
and $f$. The weak learner is nothing else that the assignment - at time $t$
- to the realization $x_{i}$ of an hypothetical proxy $h_{t}$ for $f$, whose
error is
\begin{equation}
\epsilon _{t}=\sum\limits_{i=1}^{m}D_{i}^{(t)}{I}_{[h_{t}(x_{i})\neq y_{i}]},
\end{equation}%
where ${I}$ is again the indicator function. AdaBoost assigns to $h_{t}$ a
weight inversely proportional to the error of $h_{t}$:
\begin{equation*}
w_{t}=\frac{1}{2}\log (\frac{1}{\epsilon _{t}}-1).
\end{equation*}
At the end of round $t$, AdaBoost updates the distribution so that the
samples on which $h_{t}$ errs will get a higher probability mass while
samples on which $h_{t}$ is correct will get a lower probability mass. This
will force the weak learner to focus on the problematic samples in the next
round. The output of the AdaBoost algorithm is a "strong" classifier that is
based on a weighted sum of all the weak hypotheses.

\subsubsection{Single-default performance evaluation metrics}

Performance evaluation depends on the comparison between the predictions of
any classification model and the actual realizations observed in the data.
Define the cases of single obligor default as positives and non-default as
negatives. The possible outcomes in the dataset are then true positives (%
\textit{TP}) if defaulted customers have been classified as defaulters, or
predicted to default by the model. True negatives (\textit{TN}) are
non-defaulted customers that have been predicted not to default. False
positives (\textit{FP}) are non-defaulted customers that have been predicted
to default, and false negatives (\textit{FN}) are defaulted customers that
have been predicted not to default.

A very  common metric in a classification task, which serves to highlight
the likelihood of a predicted default to be a true one, is the precision. It
counts the number of correct positive predictions, or true defaults, over
the total number of positive predictions, or predicted defaults, namely true
positives plus false positives:
\begin{equation}
Precision=\frac{TP}{TP+FP}.
\end{equation}

Another measure relevant in credit risk, which  points to the likelihood of
a correct prediction to be a default,is the sensitivity or recall. It counts
the true positive occurrences, or defaults which are predicted and do occur,
over the total number of true predictions.\footnote{%
Other common metrics, which we do not compute below are accuracy and
specificity. Accuracy is n omnibus metric, independently of the
classification. It is defined as the ratio of the total number of correct
classifications to the total number of observations:
\begin{equation}
Accuracy=\frac{TP+TN}{TP+TN+FP+FN}.
\end{equation}%
\par
A metric that is more relevant in credit risk, where false positives are the
result of caution, while true negatives should be as precise as possible, is
specificity. It is defined as the proportion of non-defaulted customers who
are correctly identified by the model as not likely to default:
\begin{equation}
Specificity=\frac{TN}{FP+TN}.
\end{equation}%
}
\begin{equation}
Sensitivity=\frac{TP}{TP+TN}.
\end{equation}

Since sensitivity and precision are of equal importance for our analysis,
one can consider as well the F1-score, which is is the  harmonic weighted
average of precision and sensitivity. The definition of the F1-score is
\begin{equation}
\begin{aligned} F1&=(1+\beta^2)\frac{Precision \, \cdot\,
Sensitivity}{Sensitivity+\,\beta^2\,\cdot\,Precision}\\
&=\frac{(1+\beta^2)TP}{(1+\beta^2)TP+\beta^2 FN+FP}, \end{aligned}
\end{equation}%
where $\beta $ is a weight parameter. Since in our discussion both measures
of precision and sensitivity are equally relevant, the weight is set to $%
\beta =1$. \newline
A further way to evaluate the results from different classification models is
to analyse the Receiver Operating Characteristic (ROC) curve and its Area
Under the Curve (AUC). The curve obtains by plotting on the horizontal axis
the number of predicted defaults which turned out in survivals - or false
positive - over the total actual survivals, and on the vertical the recall.
The ROC curve is a monotone increasing function mapping $(0,1)$ to $(0,1)$.
An uninformative test is one such that $Sensitivity(c)$ = $(1-Specificity)(c)
$ for every $c$ in $(0,1)$. This situation is represented by a ROC curve
equal to the bisector of the square $(0,1)\times (0,1)$. The most important
numerical index used to describe the behavior of the ROC curve is therefore
the area under the ROC curve (AUC). The closer this area to one, the better
the overall performance of the model, because the closer to zero is the
number of false positives, and the closer to one is the number of true
positives.

\section{Case study}

\label{DD}

This section illustrates the credit card data and its basic features,
together with the robustness procedures applied to it for probability fitting.

{We use the Kaggle database on credit card defaulters,\footnote{%
https://www.kaggle.com/uciml/default-of-credit-card-clients-dataset} which
is a collection of data from 30000 clients of a bank issuing credit cards in
Taiwan, from April to September 2005. Both the descriptive analysis of this
Section and most of the results obtained in the next Section are obtained
using the Scikit learn library in Python. For each client the dataset
contains the values of $m=24$ covariates $\boldsymbol{X}$, which are listed
and described in Appendix \ref{covar}. They capture both socio-economic
features of the client and her financial relationship with the card issuer
(repayment status, bill amount etc.). Some covariates, such as the monthly
repayment status, the past payments, the past bill amounts, are lagged
values of the same variable. The {variable $\boldsymbol{Y}$ represents the
occurrence or not of default in the next period of time. It takes the value 1
if default occurs, 0 otherwise. }Descriptive statistics of this dataset are
in Appendix \ref{covar}. }

{As one can easily argue, because the predicted event is default, and credit
card clients are chosen so as to have a minimum credit standing, the dataset
is unbalanced, with a percentage of non defaulters ({$77.88\%)$} higher  than
the one of defaulters ({$22.12\%).$ } The unbalanced nature of the dataset is
not extreme and we do to not correct it. However, we  implement different
oversampling techniques for the minority class of Y (SMOTE and Random
Oversampling) to study the robustness of such choice. The oversampling
techniques do not significantly improve the performance of the model.}

Once we have decided not to balance the dataset, we first study whether there is
multicollinearity among the covariates, then how to split it into a training
and a test set, and we perform robustness tests of the decision taken about
handling multicollinearity and the split. Aside, when predicting via the ML
approaches,  we calibrate the ML probabilities.

{To investigate whether there is multicollinearity among the covariates $%
\boldsymbol{X}$, we build their correlation matrix, which is presented in
Appendix \ref{covar}, as Figure \ref{corr}. The correlation matrix shows
that the variables representing the payment status in the previous and
current period, as well as the variables representing the bill amount up to
6 months before and after the current date, are highly correlated.
Nevertheless,we do not remove them. As a robustness test of the validity of
our decision, we  implement a feature selection via the Variance Inflation
Factor method (VIF). The overall performance of the models worsens with
feature selection. Note that in the LR model  multicollinearity does
affects the magnitude of the regression coefficients, that are not of
interest to our purposes.}

{\ Even combining the techniques of  balancing the  dataset and feature
selection does not yield better results than those presented and used in
this study. }

We then {divide the dataset into a train set and a test one. The criterion
adopted to divide the dataset is the following: 2/3 belong to the training
set and 1/3 to the test. As a consequence, our training set contains 24000
observations of the 24 covariates $\boldsymbol{X}$ and the outcome $Y$,
while the test one contains 6000 of them, $n=6000$. }

In order to fine-tune the parameter set of each model considered and
thus obtain a robust model, a Grid Search k Fold Cross Validation was
implemented on the training dataset with k = 5 folds.
We calibrate the probabilities produced by the ML methods. {By calibrated
probabilities we mean probabilities that reflect the actual underlying
frequencies. First  we verify through the so-called calibration curve that
the probabilities predicted by the three ML models were not well correctly
reproducing the frequencies of default and no-default observed as }$\mathbf{Y%
}${\ in the data  (%
{see Figure
\ref{calib} in Appendix \ref{covar}}). In  KNN  the tails of the
probability distribution are too heavy with respect to actual frequencies,
because they tend to push the predictions towards 0 and 1. RF produces a
slight systematic underestimation of the frequencies. AB, by pushing the
predicted probabilities away from 0 and 1, makes the center of the
theoretical distribution too heavy. As a result,  for each model we adopt a
second model, called a calibrator, capable of correcting the accurate but
uncalibrated probabilities produced by the classifier into probabilities
closer to actual frequencies. There are essentially two ways to calibrate
the probabilities predicted by a classifier: Platt Scaling and Isotonic
Regression. We  choose the calibrator that minimises the Expected
Calibration Error (ECE). We apply this procedure to the three Machine
Learning models, leaving the probabilities predicted by the Logistic
Regression unchanged,  to avoid overfitting. The probabilities predicted by
the calibrators that minimised the ECE are those given by the Isotonic
regression for AdaBoost and Random Forest,  those given by Platt Scaling for
KNN, see Table \ref{CERR}. }

\begin{table}[h]
\begin{center}
\begin{tabular}{|l|l|l|l|l|}
\hline
Model & Calibration method & expected calibration error   \\ \hline
RF & Isotonic regression & 2.38\%  \\ \hline
AB & Isotonic regression & 1.79\%   \\ \hline
KNN & Platt & 5.65\%   \\ \hline
\end{tabular}
\end{center}
\caption{Calibration errors.}
\label{CERR}
\end{table}
{We use the calibrated  proabilities for  joint default modeling.}

\section{Results}

\label{Res} This Section  evaluates the performance of each $h=$ LR, RF, AB,
KNN and reports the coefficients for the estimates of  the single and joint
defult probabilities.

\subsection{Performance measures}

For each of the candidate models we report in Table \ref{scores} the performance
measures on single obligors. Precision is highest for AB, which is literally
the most precise method to predict default on this database. Recall is
highest for AB, meaning that the latter is also the method which maximizes
the true defaults, considering all true predictions, on this dataset. As a
synthetic measure, consider the F1-score, which is again highest for AB. All
the three indicators are slightly lower for LR.

\begin{table}[h!]
\begin{center}
\begin{tabular}{|l|l|l|l|l|}
\hline
Model & Precision & Recall & F1-score & AUC \\ \hline
LR & 0.61 & 0.78 & 0.68 & 0.64 \\ \hline
RF & 0.79 & 0.81 & 0.78 & 0.77 \\ \hline
AB & 0.80 & 0.82 & 0.79 & 0.77 \\ \hline
KNN & 0.78 & 0.81 & 0.78 & 0.72 \\ \hline
\end{tabular}%
\end{center}
\caption{Performance measure for each model}
\label{scores}
\end{table}

\begin{figure}[t]
\centering
\includegraphics[width=0.6\linewidth]{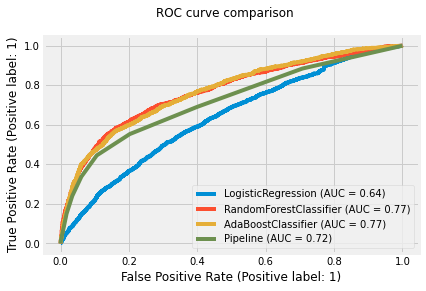}
\caption{ROC curves}
\label{ROC}
\end{figure}

To get an overall grasp of the performance of the four methods to estimate
individual default probabilities, we consider in Figure \ref{ROC} the ROC\
curve and in Table \ref{scores} last column  the area under it, the AUC, which is approximately the same for RF
and AB.

{Comparing the indices in Table \ref{scores} we conclude that the ML methods
perform better that the traditional LR, at least on this database and for
single defaults. It is therefore important to compare the discrepancy in
joint defaults that they entail. }

\subsection{Moments and mixing distributions}

\label{BMV}

\bigskip

We provide the estimates of  the moments,  relevant  for the joint defaults
and for the number of defaults, and then the parameters of the beta mixing
variables.

Table \ref{moments} provides the first two empirical moments of $Q_{h}$  and
the resulting default correlations, for LR and for each ML model.

\begin{table}[h]
\begin{center}
\begin{tabular}{|l|l|l|l|l|}
\hline
Moments & LR & RF & AB & KNN \\ \hline
$p$ & 0.2635 & 0.2232 & 0.2234 & 0.2209 \\ \hline
$\pi_2$ & 0.0883 & 0.0925 & 0.0901 & 0.1023 \\ \hline
$\rho$ & 0.0975 & 0.2462 & 0.2319 & 0.3108 \\ \hline
\end{tabular}%
\end{center}
\caption{Moments estimated with LR and with different ML techniques.}
\label{moments}
\end{table}

The LR method seems to overestimate the first and underestimate the second
order moment, if compared with the ML methods. As a consequence of the
relationship among  $p$, $\pi _{2}$ and $\rho $, ML techniques are capable
to capture higher correlations $\rho $ among default. This preliminary result has to be
further investigated to see whether is $p$ or $\rho $ that mainly affects
the tail of the loss.

As for the beta mixing variables, using the method of moments, we get the
following densities

\begin{figure}[h!]
\centering
\subfigure[AB density]{\includegraphics[scale=0.3]{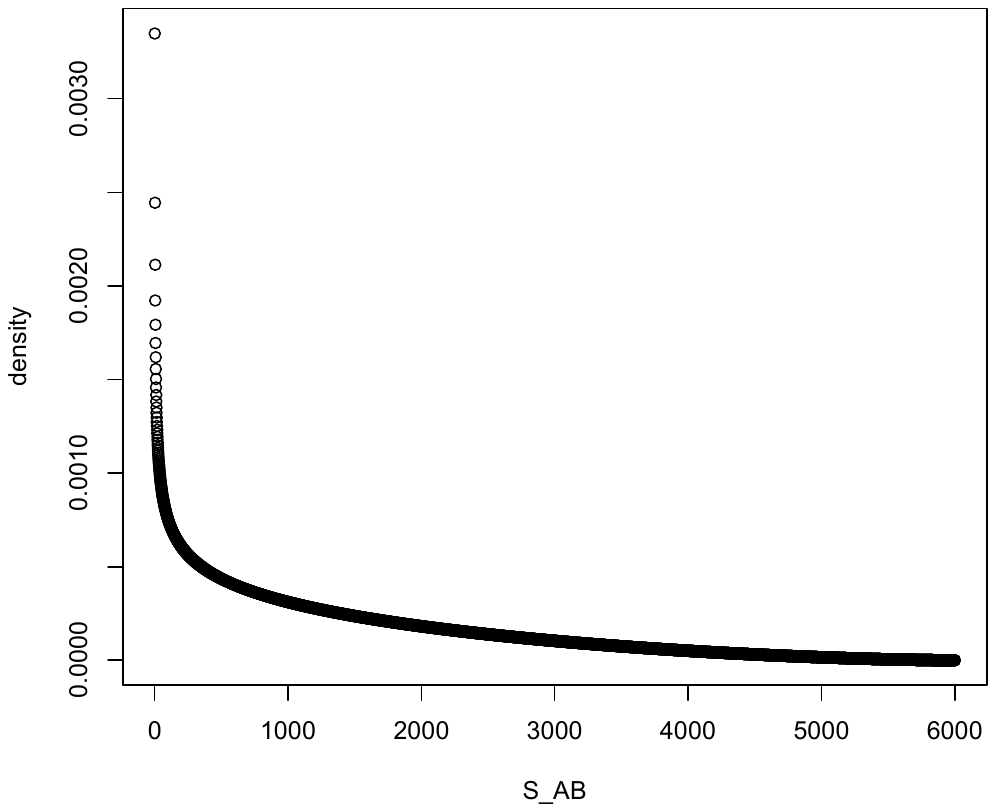}}
\subfigure[LR
density]{\includegraphics[scale=0.3]{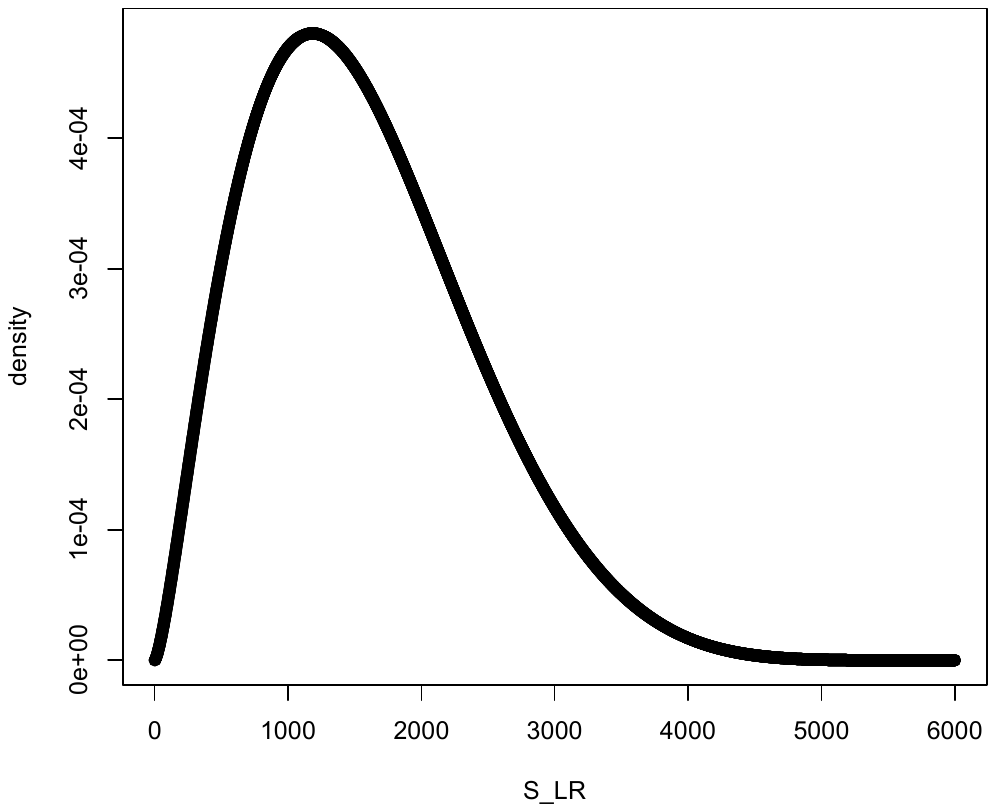}}\newline
\subfigure[RF density]{\includegraphics[scale=0.3]{ABden.pdf}} %
\subfigure[KNN density]{\includegraphics[scale=0.3]{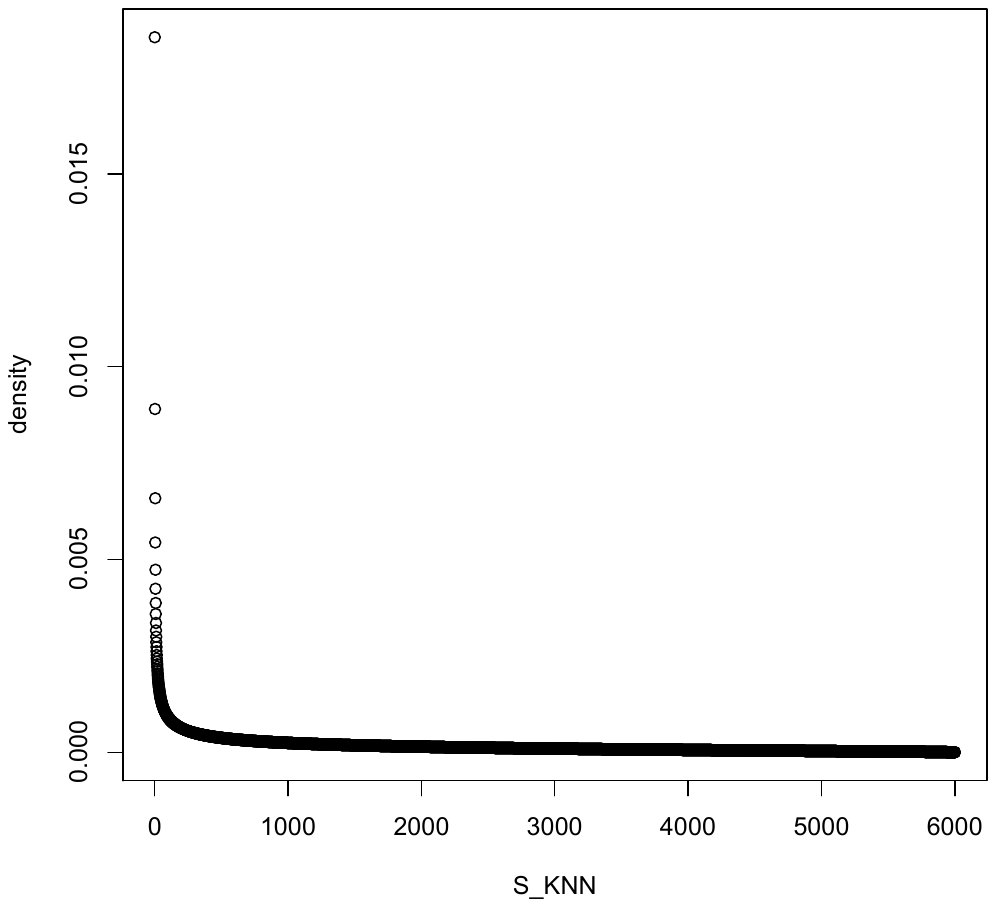}}
\caption{Beta binomial pmf estimated with LR, AB RF and KNN.}
\label{densiti}
\end{figure}
Figure \ref{densiti} shows  that the three ML methods provide similar
densities, while the LR is significantly different, also in the tails. This
will be reflected in the analysis below, particularly in the risk measures.

The corresponding parameters $a, b$ are  in Table \ref{ModMom}. The remaining
entries of the Table are the theoretical or paramerical mean and
correlation, which are the counterparts of the emprical - or non parametric
- ones in the previous Table:



\begin{table}[h]
\begin{center}
\begin{tabular}{|l|l|l|l|l|}
\hline
$\beta$-parameters & LR & RF & AB & KNN \\ \hline
$a$ & 2.42 & 0.68 & 0.73 & 0.48 \\ \hline
$b$ & 6.78 & 2.38 & 2.57 & 1.72 \\ \hline
$p$ & 0.2630 & 0.2222 & 0.2212 & 0.2182 \\ \hline
$\rho$ & 0.0980 & 0.2463 & 0.2326 & 0.3125 \\ \hline
\end{tabular}%
\end{center}
\caption{Estimated parameters of the beta distribution for each LR/ML\ model}
\label{ModMom}
\end{table}

The Kolmogorov-Smirnov test of the four $\beta $ distributions reject their
appropriateness to describe the mixing variable, since the $p$-value is
close to zero in all cases. However, Table \ref{ModMom} shows that the
theoretical moments are very close to the non parametrical ones. This means
that, despite the misspecifications for the distribution of $Q$, we expect
 theoretical and empirical $\VaR$ to be close, with a possible more
significant discrepancy in correspondence to $\alpha =0.99$ (see Section \ref%
{not}).

\bigskip

\bigskip


%
%
%
%
%

\section{$\VaR$  bounds and estimates}

\label{VaR}

Risk measurement of the credit losses is done through their $\VaR$ and $\ES$,
computed along Definition 4.1. Both on a small portfolio ($d=25$) and on the
overall portfolio ($d=6000$), we compute the $\VaR$ and the corresponding $\ES$
bounds and values at the levels of confidence $\alpha =0.9;0.95;0.99$.

\subsection{Small portfolio}

Table \ref{VarBLRs} provides the analytical bounds for the VaR, in the class
of all credit portfolio with given marginal default probability $p$,  for
each model specification. The minimum VaR is slightly higher for LR, while
all the ML techniques perform similarly. This points to the importance of $p$
in determining the aggregate loss, and  is a consequence of the fact that LR
overestimates the marginal default probability with respect to ML. It needs
to be confirmed for the high dimension portfolio, where the  differences can
be more evident.

\begin{table}[tbp]
\begin{center}
\begin{tabular}{|l|l|l|l|}
\hline
VaR & $\alpha=0.9$ & $\alpha=0.95$ & $\alpha=0.99$  \\ \hline
min LR & 5 & 6 & 7   \\ \hline
max LR & 25 & 25 & 25   \\ \hline\hline
min RF & 4 & 5 & 6  \\ \hline
max RF & 25 & 25 & 25  \\ \hline\hline
min KNN & 4 & 4 & 5 \\ \hline
max KNN & 25 & 25 & 25   \\ \hline\hline
min AB & 5 & 6 & 6   \\ \hline
max AB & 25 & 25 & 25 \\ \hline
\end{tabular}%
\end{center}
\caption{ VaR bounds for a portfolio of $n=25$ obligors, with the  $p=$
vaues of Table 4.}
\label{VarBLRs}
\end{table}

We now consider the effect of the second order and higher order moments on
joint defaults. The $\VaR$s and ESs computed with the four beta binomial or
parametrical models are reported in Table \ref{VaRs} that account only for the first and second moments, because the beta parameters do (see Appendix \ref{beta}). Table \ref{VaRs} also
reports the $\VaR$ and $\ES$ computed with the empirical distribution of $S$,
that account for sample higher order moments.%

\begin{table}[h]
\begin{center}
{\tiny
\begin{tabular}{|l|l|l|l|l|l|l|l|l|}
\hline
& LR\_e & LR\_m & RF\_e & RF\_m & AB\_e & AB\_m & KNN\_e & KNN\_m \\ \hline
$\alpha$ & VaR & VaR & VaR & VaR & VaR & VaR & VaR & VaR \\ \hline
0.99\% & 15 & 17 & 23 & 21 & 22 & 21 & 24 & 23 \\ \hline
0.95\% & 13 & 14 & 19 & 17 & 18 & 16 & 20 & 18 \\ \hline
0.90\% & 12 & 12 & 15 & 14 & 15 & 14 & 16 & 15 \\ \hline
$\alpha$ & ES & ES & ES & ES & ES & ES & ES & ES \\ \hline
0.99\% & 18.5275 & 17.5120 & 23 & 21.0017 & 22.0002 & 21.0017 & 24 & 23 \\
\hline
0.95\% & 16.0755 & 15.5503 & 19.0084 & 17.1024 & 18.0318 & 16.2864 & 20.0019
& 18.0318 \\ \hline
0.90\% & 14.7849 & 14.7849 & 15.3527 & 14.7752 & 15.3527 & 14.7752 & 16.1432
& 15.3527 \\ \hline
\end{tabular}
}
\end{center}
\caption{Non parametrical vs beta-binomial VaRs and ESs of the distribution
of number of defaults for a portfolio of $25$ obligors. LR\_e; RF\_e ,
AB\_e, KNN\_e are the non parametrical distributions for each model and
LR\_m,RF\_m, AB\_m, KNN\_m are the parametrical distributions for each
model. }
\label{VaRs}
\end{table}
Compare then the $\VaR$s and ESs obtained for the three ML methods with
the $\VaR$ and $\ES$ obtained with LR. {The }${\VaR}${s and }{ES}{s obtained
with ML techniques are significantly higher than the ones obtained with LR,
independently of whether we look at the emprical or theoretical ones. So,
when we include correctly the covariates and obligors' dependencies, as ML by
definition does, we do not underestimate risk. }This -compared to the bounds- points to the
importance of the moments beyond   $p$ in determining the aggregate loss. It needs to be confirmed
for the high dimension portfolio, where the differences can be more evident.

{Comparing the non parametrical vs the beta $\VaR$ for each model in Table %
\ref{VaRs}, we obtain the following results. For the LR case the beta $\VaR$
is always bigger than the non parametrical one, showing that LR is not
capable to capture heavy tails. This is more evident for $\alpha =0.99$,
where the importance of the tails of $Q$ is higher. On the contrary, for all
the ML techniques the non parametrical $\VaR$ is higher than the beta one.
This is in line with the ability of these techniques to capture complex
dependencies. For all the models the non parametrical and beta $\VaR$ are
close, something which supports the idea  that the estimates of $p$ and $%
\rho $ are more important than the higher moments. Indeed, only the former
determine the beta parameters, while all the moments subsume the
non-parametrical specification of a distribution for $Q_{h}$. Similar
comments can be made for the ESs in Table \ref{VaRs}. This result cannot be
confirmed on the high dimension portfolio, since we do not use for it the
emprical distribution. }

In Table 7 we compare the fit to the non parametrical distribution of $Q_{h}$%
, $h=$LR, RF, AB, KNN, of the corresponding beta distribution, using the
Kullback-Leibler (KL) distance.
\begin{table}[h]
\begin{center}
\begin{tabular}{|l|l|l|l|l|}
\hline
& LR & RF & AB & KNN \\ \hline
KL distance & 0.0450 & 0.1153 & 0.0783 & 0.1557 \\ \hline
\end{tabular}%
\end{center}
\caption{KL distance for each model.}
\label{KL}
\end{table}
The conclusion from Table \ref{KL} is that the parametrical pmf is closer to
the non parametrical one for the AB  technique. We therefore focus on AB and
the benchmark LR to estimate the $\VaR$ and $\ES$ for the large portfolio.

\subsection{Large portfolio}

We now consider a large portfolio, $d=n=6000$, where the differences between
LR and ML are more evident than in the small portfolio, as expected.

\textcolor{black}{Table \ref{VarBLR} provides the analytical bounds for the VaR, in the class of all credit portfolios with given marginal default probability $p$, estimated respectively with AB and LR. As one can see, since LR overestimates $p$ with respect to $AB$, without specifying the correlation among defaults, the minimum VaR is higher for LR.
} This confirms what we concluded about $p$ - and the importance not to
overestimate it - from a small portfolio.
\begin{table}[tbp]
\begin{center}
\begin{tabular}{|l|l|l|l|l|}
\hline
VaR & $\alpha=0.9$ & $\alpha=0.95$ & $\alpha=0.99$ \\ \hline
min AB & 823 & 1096 & 1294  \\ \hline
max AB & 6000 & 6000 & 6000   \\ \hline\hline
min LR & 1091 & 1384 & 1573 \\ \hline
max LR & 6000 & 6000 & 6000   \\ \hline
\end{tabular}%
\end{center}
\caption{ VaR bounds, for the $p$ values of Table 4.}
\label{VarBLR}
\end{table}

\begin{figure}[h!]
\centering
\includegraphics[scale=0.3]{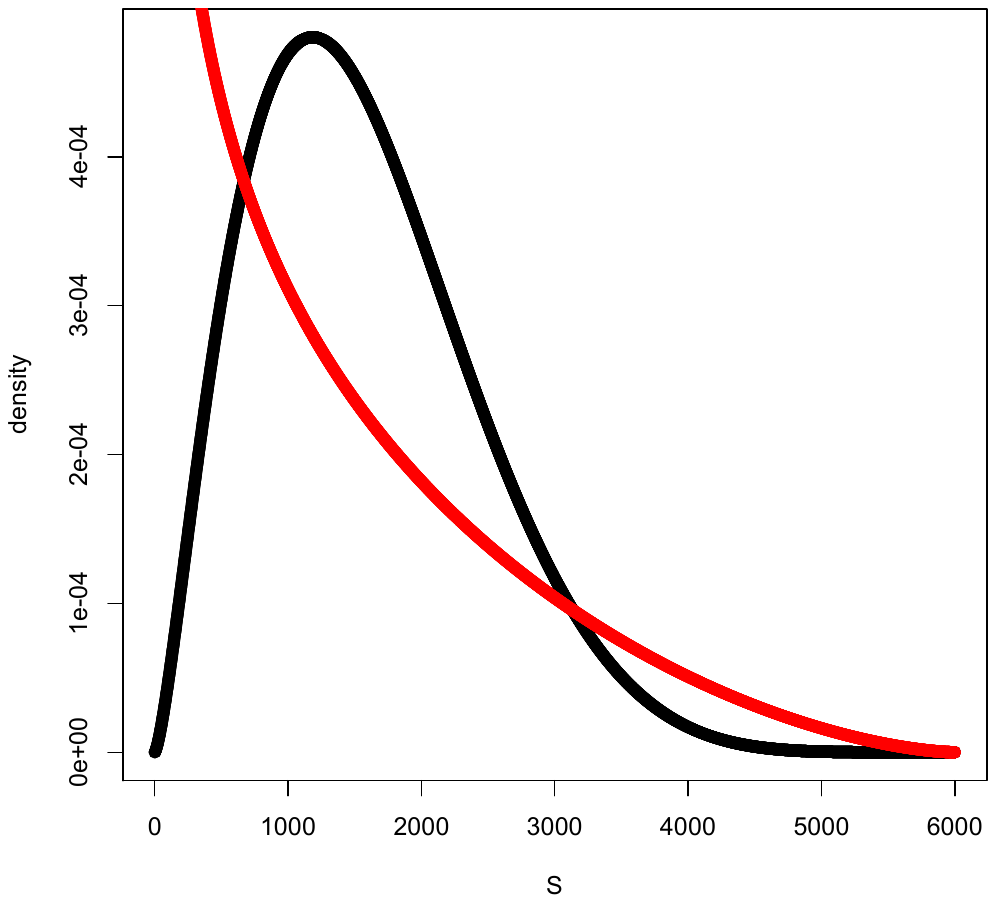}
\caption{Beta binomial pmf estimated with LR (black) and AB (red).}
\label{AR}
\end{figure}

We now analyse the effect of dependence on the portfolio risk. Our aim is to
stress the differences between ML and LR.  We
compare the portfolio estimated with AB with our benchmark, LR. In Figure %
\ref{AR}, the pmf from AB and LR are overlapped to exhibit the differences
in the tail distribution.

The theoretical $\VaR$s and $\ES$s obtained from the beta-binomial models
corresponding to the parameter $a$ and $b$ estimated using LR and AB and
with $d=6000$ are reported in Table \ref{VarSim}. Computations are made with
the support of $R$.

\textcolor{red}{\begin{table}[h!]
\begin{center}
\begin{tabular}{|l|l|l|l|l|}
\hline
$\alpha$ & VaR LR & VaR AB &ES LR& ES AB  \\ \hline
0.99\% & 3794 & 4798&4099.6&5131.9 \\ \hline
0.95\% & 3107 & 3788&3524.3&4395.1  \\ \hline
0.90\% & 2729 & 3139&3213.0&3916.9   \\ \hline
\end{tabular}\end{center}
\caption{Beta-binomial VaR and ES for large portfolios.}
\label{VarSim}
\end{table}} \ Both VaRs and ESs show that using ML to classify
individual defaults we have significantly higher measures of risk. The
difference between the LR and AB VaRs and $ESs$ in Table \ref{VarSim} is
higher for the quantile $\alpha =0.99$. The Table confirms what we concluded
about the dependence - and the importance not to underestimate it - from a
small portfolio.

We thus have empirical evidence of the importance of ML techniques to
properly specify a credit risk model, since they allow us to incorporate the risk
coming from dependencies among covariates, that seems to be the more relevant
factor affecting the tail of the loss, even without using all moments.

%
%
%
%


%





\newpage

\appendix

\section{Beta and Beta Binomial distributions}

\label{beta}

Let $Q$ have a Beta distribution of parameters $a$ and $b$, i.e. $Q\sim
\beta(a,b)$, its density $g_Q$ is given by
\begin{equation}
g_Q(q)=\frac{1}{\beta(a,b)}q^{a-1}(1-q)^{b-1}, \qquad a,b>0, \quad 0<q<1
\end{equation}
where
\begin{equation}
\beta(a,b)=\int_{0}^{1}z^{a-1}(1-z)^{b-1}dz, \qquad 0<z<1
\end{equation}
Standard calculations give:
\begin{equation}
\pi_k = \prod_{j=0}^{k-1}\frac{a+j}{a+b+j},
\end{equation}
in particular
\begin{equation*}
\pi=\frac{a}{a+b},\,\,\, \text{and}\,\,\,\rho_y =\frac{1}{a+b+1}.
\end{equation*}%
\newline
A Beta Binomial random variable $S$ with parameters $a,b$ and $d$, $S\sim
\beta Bin(a,b,d)$, has pmf $p_S(k)$ given by
\begin{equation}  \label{betbin}
p_S(k)=\mathbb{P}(S=k)= \binom{d}{k}\frac{\beta(a+k,b+d-k)}{\beta(a,b)},
\,\,\, k\leq d.
\end{equation}

\section{Dataset description}

\label{covar}

For each client the dataset contains the values of 24 covariates, which
include demographic variables, repayment status, past payments, past bill
amount, and an indicator of default, namely:

\begin{itemize}
\item ID: ID of each client

\item LIMIT\_BAL: Amount of the given credit (NT dollar), which includes
both the individual consumer credit and his/her family (supplementary)
credit.

\item SEX: Gender (1 = male; 2 = female)

\item EDUCATION: (1 = graduate school; 2 = university; 3= high school; 4=
others; 5= unknown; 6= unknown)

\item MARRIAGE: Marital status (1 = married; 2 = single; 3 = others)

\item AGE: Age in years

\item PAY\_1 to 6 (with 1 missing): Repayment status in September to April
2005 (-1=pay duly, 1=payment delay for one month, 2=payment delay for two
months, \ldots\ 8=payment delay for eight months, 9=payment delay for nine
months and above)

\item BILL\_AMT1 to 6: Amount of bill statement in September to April, 2005
(NT dollar)

\item PAY\_AMT1: Amount of previous payment in September to April, 2005 (NT
dollar)

\item default.payment.next.month: Default payment (1=yes, 0=no)
\end{itemize}

Table \ref{Stat} provides the descriptive statistics of the data.

\begin{table}[!ht]
\centering
\resizebox{0.9\textwidth }{!}{
    \begin{tabular}{|l|l|l|l|l|l|l|l|}
    \hline
        - & ID & LIMIT\_BAL & SEX & EDUCATION & MARRIAGE & AGE & - \\ \hline
        count & 30000 & 30000 & 30000 & 30000 & 30000 & 30000 & - \\ \hline
        mean & 15000.5 & 167484.3227 & - & - & - & 35.4855 & - \\ \hline
        std & 8660.398374 & 129747.6616 & - & - & - & 9.217904068 & - \\ \hline
        min & 1 & 10000 & 1 & 0 & 0 & 21 & - \\ \hline
        25\% & 7500.75 & 50000 & 1 & 1 & 1 & 28 & - \\ \hline
        50\% & 15000.5 & 140000 & 2 & 2 & 2 & 34 & - \\ \hline
        75\% & 22500.25 & 240000 & 2 & 2 & 2 & 41 & - \\ \hline
        max & 30000 & 1000000 & 2 & 6 & 3 & 79 & - \\ \hline
        - & BILL\_AMT1 & BILL\_AMT2 & BILL\_AMT3 & BILL\_AMT4 & BILL\_AMT5 & BILL\_AMT6 & - \\ \hline
        count & 30000 & 30000 & 30000 & 30000 & 30000 & 30000 & - \\ \hline
        mean & 51223.3309 & 49179.07517 & 47013.1548 & 43262.94897 & 40311.40097 & 38871.7604 & - \\ \hline
        std & 73635.86058 & 71173.76878 & 69349.38743 & 64332.85613 & 60797.15577 & 59554.10754 & - \\ \hline
        min & -165580 & -69777 & -157264 & -170000 & -81334 & -339603 & - \\ \hline
        25\% & 3558.75 & 2984.75 & 2666.25 & 2326.75 & 1763 & 1256 & - \\ \hline
        50\% & 22381.5 & 21200 & 20088.5 & 19052 & 18104.5 & 17071 & - \\ \hline
        75\% & 67091 & 64006.25 & 60164.75 & 54506 & 50190.5 & 49198.25 & - \\ \hline
        max & 964511 & 983931 & 1664089 & 891586 & 927171 & 961664 & - \\ \hline
        - & PAY\_1 & PAY\_2 & PAY\_3 & PAY\_4 & PAY\_5 & PAY\_6 & - \\ \hline
        count & 30000 & 30000 & 30000 & 30000 & 30000 & 30000 & - \\ \hline
        mean & - & - & - & - & - & - & - \\ \hline
        std & - & - & - & - & - & - & - \\ \hline
        min & -2 & -2 & -2 & -2 & -2 & -2 & - \\ \hline
        25\% & -1 & -1 & -1 & -1 & -1 & -1 & - \\ \hline
        50\% & 0 & 0 & 0 & 0 & 0 & 0 & - \\ \hline
        75\% & 0 & 0 & 0 & 0 & 0 & 0 & - \\ \hline
        max & 8 & 8 & 8 & 8 & 8 & 8 & - \\ \hline
        - & PAY\_AMT1 & PAY\_AMT2 & PAY\_AMT3 & PAY\_AMT4 & PAY\_AMT5 & PAY\_AMT6 & def\_pay \\ \hline
        count & 30000 & 30000 & 30000 & 30000 & 30000 & 30000 & 30000 \\ \hline
        mean & 5663.5805 & 5921.1635 & 5225.6815 & 4826.076867 & 4799.387633 & 5215.502567 & 0.2212 \\ \hline
        std & 16563.28035 & 23040.8704 & 17606.96147 & 15666.15974 & 15278.30568 & 17777.46578 & 0.415061806 \\ \hline
        min & 0 & 0 & 0 & 0 & 0 & 0 & 0 \\ \hline
        25\% & 1000 & 833 & 390 & 296 & 252.5 & 117.75 & 0 \\ \hline
        50\% & 2100 & 2009 & 1800 & 1500 & 1500 & 1500 & 0 \\ \hline
        75\% & 5006 & 5000 & 4505 & 4013.25 & 4031.5 & 4000 & 0 \\ \hline
        max & 873552 & 1684259 & 896040 & 621000 & 426529 & 528666 & 1 \\ \hline
    \end{tabular}}
\caption{Descriptive statistics.}
\label{Stat}
\end{table}

The correlation matrix of the 24 covariates is presented here as Figure \ref%
{corr}.
\begin{figure}[t]
\centering
\includegraphics[width=0.7\linewidth]{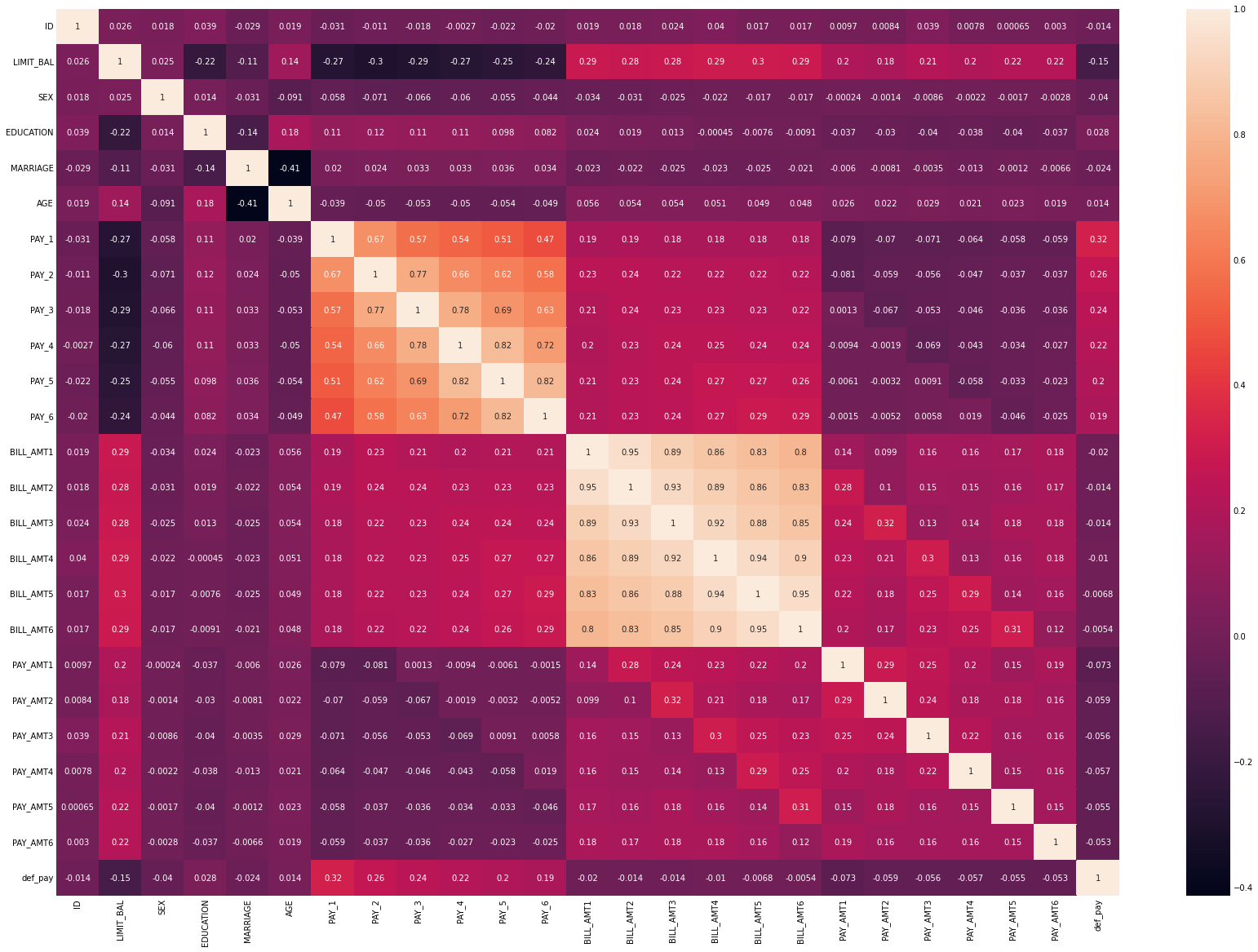}
\caption{Correlation matrix of the variables.}
\label{corr}
\end{figure}

\begin{figure}[h!]
\centering
\subfigure[AB  curve]{\includegraphics[scale=0.3]{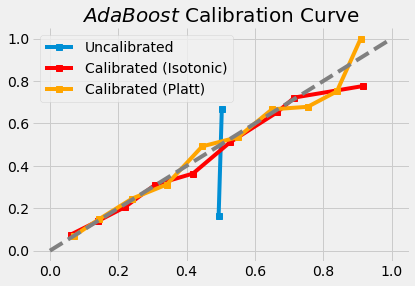}} %
\subfigure[RF  curve]{\includegraphics[scale=0.3]{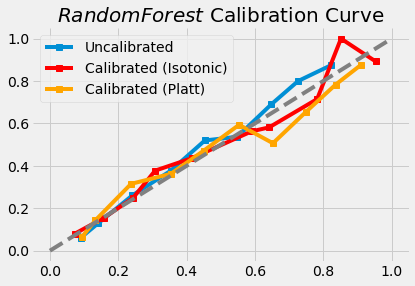}} %
\subfigure[KNN  curve]{\includegraphics[scale=0.3]{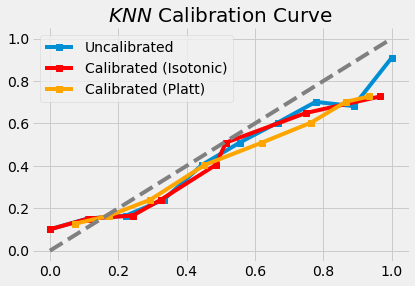}}
\caption{Calibration curves for  AB RF and KNN.}
\label{calib}
\end{figure}

Figure \ref{calib} provides the calibration curves of default probabilities
for each ML method.

\section*{Acknowledgements}

Elisa Luciano and Patrizia Semeraro gratefully acknowledge financial support
from the Italian Ministry of Education, University and Research, MIUR,
"Dipartimenti di Eccellenza" grant 2018-2022. (E11G18000350001). Patrizia
Semeraro also thanks Gianluca Mastrantonio for helpful discussions.

\newpage 
\bibliographystyle{apalike}
\bibliography{biblio}

\end{document}